\documentclass[10pt,twocolumn,aps,pra,amsmath,amssymb,showpacs]{revtex4-1}
\usepackage{bm}
\usepackage{mathrsfs}
\usepackage{graphicx}
\usepackage[usenames,dvipsnames,svgnames,table]{xcolor}
\usepackage[unicode=true,
            pdfusetitle, 
            bookmarks=true,
            bookmarksnumbered=false,
            bookmarksopen=false,
            breaklinks=true,
            pdfborder={0 0 0},
            backref=false,
            colorlinks=true]{hyperref}
\hypersetup{linkcolor=NavyBlue,urlcolor=NavyBlue,citecolor=NavyBlue}

\usepackage{amsthm}

\providecommand{\propositionname}{Proposition}

\usepackage{tikz-cd}

\newcommand{\cH}{\mathcal{H}}

\begin{document}

\title{Entanglement increase from local interaction in the absence of initial quantum correlation in the environment and between the system and the environment}
\author{Iman Sargolzahi}
\email{sargolzahi@neyshabur.ac.ir, sargolzahi@gmail.com}
\affiliation{Department of Physics, University of Neyshabur, Neyshabur, Iran}
\author{Sayyed Yahya Mirafzali}
\email{y.mirafzali@vru.ac.ir}
\affiliation{Department of Physics, Faculty of Science, Vali-e-Asr University of Rafsanjan, Rafsanjan, Iran}

\begin{abstract}
We consider a bipartite quantum system $S=AB$ such that the part $A$ is isolated from the environment $E$ and only the part $B$ interacts with $E$.
Under such circumstances, entanglement of the system may experience decreases and increases, during the evolution of the system.
 Here, we show that the entanglement of the system can exceed its initial value, under such local interaction, even though, at the initial moment, there is no entanglement in the environment and the system and the environment are only classically correlated.
The case which is studied in this paper possesses another interesting feature too:  The reduced dynamics of the system can be modeled as a completely positive map.
In addition, we introduce the concept of \textit{inaccessible entanglement} to  explain why entanglement can exceed its initial value, under local interactions, in open quantum systems. 
 \end{abstract}

%\pacs{03.65.Yz, 03.67.-a}

\maketitle
%Introduction
\section{Introduction}
%Introduction
Recently, many studies have been focused on the dynamics of entanglement in open quantum systems, both in bipartite and multipartite cases \cite{1}. Entanglement may decrease or even experience revivals during the interaction of the system with the environment. 

An important case is when the system is bipartite and each part interacts with its local (quantum) environment \cite{2, 2aa}. 
One may expect that in this case only entanglement decrease (sudden death) will occur. 
%, since entanglement can not increase under local operations. 
But, interestingly,  entanglement revivals occur in such cases \cite{2, 2a, 3, 4, 10,  11}. This phenomenon is usually explained as a consequence of the non-Markovianity of the dynamics and so the memory effects of the environment. 
%In addition, as we will see in Sec.~\ref{sec:inaccessible}, this phenomenon does not violate the LOCC paradigm.

A more interesting case is when the entanglement of the system exceeds its initial value.
%In this paper, we focus on the following question: Can entanglement, in an open bipartite system, when  each part interacts with its local environment, exceed its initial value?
 An interaction (with the environment) which is local according to a bipartition of the system, can be nonlocal according to another bipartition and so leads to entanglement increase (according to this latter bipartition) in the system \cite{11b}. 
% But, one can also find examples for which entanglement exceeds its initial value, under local interaction (according to the bipartition for which the interaction is local).
 Also, if the 
 the evolution of one part of our bipartite system is given by a non-Hermitian Hamiltonian, then  the entanglement of the  system can exceed its initial value \cite{9}.

Interestingly, even in the context of the conventional quantum mechanics,  one can also  find  examples for which entanglement exceeding,  under local interactions, occurs  \cite{8, 2b, 6, 7,  11a, 5}. In such cases, though each part of our bipartite system interacts with its local environment,  the reduced dynamics of the system is not given by local operations and so, entanglement exceeding occurs in the system. In other words, though the  dynamics of the whole system-environment is given by a local operation (a local completely positive map), the reduced dynamics of the system is not so and this can lead to entanglement increase.
% For example, in Ref. \cite{6}, a bipartite configuration has been considered, such that, in each part, there is a single cavity mode and an $N$-mode reservoir ($N\rightarrow\infty$),  interacting with each other. Then, it has been shown that the entanglement between the two reservoirs, which is zero initially, appears suddenly, at a time $t_{1} >0$. This phenomenon is called \textit{entanglement sudden birth} \cite{6}.

Now, an important question arises: If the entanglement of the system starts to increase at time $t_{1}$ and increases monotonically during the time interval $\left[ t_{1}, t_{2} \right]$  ($t_{1}< t_{2}$), 
can the reduced dynamics of the system, from $t_{1}$ to $t_{1}^{\prime}$, where $t_{1}^{\prime}\in \left( t_{1}, t_{2} \right]$, be given by a (nonlocal) completely positive map?

This question has been considered in Ref. \cite{8}. It has been shown that, for the case studied there, the time evolution of the system, from $t_{1}$ to $t_{1}^{\prime}$, is given by a non-completely-positive map, in fact, by a non-trace-preserving map.

Interestingly, entanglement  exceeding in the system can occur even  when the reduced dynamics of the system is completely positive. For example, in Ref. \cite{2b}, a four-qubit case $AE_{A};BE_{B}$  has been considered, where $E_{A}$ and $E_{B}$ are two separated atoms, each interacting with its local cavity mode, $A$ and $B$, respectively. The initial state of   $AE_{A};BE_{B}$  was chosen as  
\begin{equation}
\label{eq:1a}
\vert\Phi_{0}\rangle= \vert 0_{A}\rangle\otimes\vert 0_{B}\rangle\otimes \vert\Psi_{E_{A}E_{B}}\rangle,
\end{equation}
where $\vert\Psi_{E_{A}E_{B}}\rangle$ is an entangled state in $E_{A}E_{B}$ and $\vert 0_{A}\rangle$ and $\vert 0_{B}\rangle$ are some fixed states (the vacuum states) in $A$ and $B$, respectively. Let's consider $AB$ as our bipartite system and $E_{A}$ and $E_{B}$ as local environments of $A$ and $B$, respectively. So, the entanglement of the system $AB$, which is initially zero, can increase just from the initial moment $t=0$  \cite{2b}. The initial state of the system-environment in Eq. (\ref{eq:1a}) is factorized. So, the reduced dynamics of the system is completely positive \cite{20}. Therefore, entanglement increase can occur even when the reduced dynamics of the system is completely positive.

Note that the initial state of the environment $E_{A}E_{B}$ in Eq. (\ref{eq:1a})  is entangled. So, one can argue that the transfer of  the entanglement from the environment to the system results in entanglement exceeding in the system

Can we find a case in which entanglement exceeding occurs in the system, even when the environment is initially unentangled and the reduced dynamics of the system is, in addition, completely positive? Finding such a case is the subject of this paper.

In this paper, we consider a two-qubit system $S=AB$, such that the qubit $A$ is isolated from the environment and only the qubit $B$ is interacting with the environment $E$.
In our case:
% at the initial moment, there is no entanglement in the environment and, in addition, the system and the environment are only classically correlated. In spite of these properties, the entanglement of the system exceeds its initial value, just from the initial moment $t=0$. 
% In addition,  the reduced dynamics of the system is given by a completely positive map.
% 
% In other words, the main result of this paper is that entanglement exceeding occurs even when we possess the three following features,  simultaneously:
 
 (1) Unlike the cases studied in Refs. \cite{2b, 6, 7, 5},  there is no (initial) entanglement in the environment and so, there is no transfer of the entanglement from the environment to the system, during the evolution.
 
 (2) Unlike (the first initial state considered in) Ref. \cite{11a}, there is no initial entanglement between the subsystem $A$ ($B$) and the environment $E$ and so, entanglement exceeding in the system $S=AB$ cannot be related to the transfer of this initial entanglement from $AE$ ($BE$) to $AB$.
 
 (3) And, finally, (unlike  Ref. \cite{8})  the reduced dynamics of the system $S=AB$ is given by a completely positive map (which is, obviously,  nonlocal, otherwise no entanglement exceeding is possible). 
 
Finding this interesting  case is, to some extent, due to our previous results in  Refs. \cite{12, 13}. Especially, we will use the following result of Ref. \cite{13}: Under local interactions, entanglement increase can occur only when the initial state of the whole system-environment is not a so-called \textit{Markov state}.

Exceeding the entanglement can not be related to the memory effects of  the environment, in general. When the entanglement of an open system initially decreases and then revives, we can say that the environment stores the entanglement of the system, during its decrease, and then gives it back to the system, during the revival. In general, the above explanation is no more valued for the case that the entanglement of the system exceeds its initial value.

In addition, as stated before, when the initial entanglement of the environment and the initial entanglement between the system and the environment are zero, entanglement exceeding in the system can not be related to the transfer of the entanglement from the environment to the system.

 In such circumstances, we argue that there is an initial supply of entanglement in the \textit{ whole system-environment}, which is initially inaccessible for the system, and the transfer of (a part of) this supply to the system, during the interaction of the system and the environment, results in exceeding the entanglement of the system.

The paper is organized as follows. In the next section, we  introduce the concurrence, an entanglement monotone which  will be used in this paper. Markov states are introduced in Sec. ~\ref{markov}. In addition, their role, in the phenomenon of entanglement exceeding,  is discussed there.  In Sec. ~\ref{sec:model}, the model system and the related results are given. Section ~\ref{sec:inaccessible} is on the explanation of the phenomenon of entanglement exceeding,  introducing the concept of \textit{inaccessible entanglement}. 
Finally, our paper is ended in Sec.~\ref{sec:summary}, with a summary of our results.

\section{Concurrence}\label{sec:Concurrence}
%Concurrence
Consider a bipartite system $S=AB$. 
For a pure  state $\vert\psi\rangle \in \cH_{A}\otimes\cH_{B}$, where $\cH_{A}$ and $\cH_{B}$ are the Hilbert spaces of the subsystems $A$ and $B$, respectively, concurrence is defined as \cite{14}
\begin{equation}
\label{eq:1}
C(\vert\psi\rangle)= \sqrt{2\left[1-\mathrm{Tr}(\rho_{r}^{2})\right]} ,
\end{equation}
where $\rho_{r}$ is the reduced state of either the subsystem $A$ or the $B$.
$C(\vert\psi\rangle)=0$ if and only if $\vert\psi\rangle$ is a product state. The generalization of the above definition for mixed states is as \cite{14}
\begin{equation}
\label{eq:2}
C(\rho)= \min_{\lbrace p_{i},\vert\psi_{i}\rangle \rbrace} \sum_{i} p_{i}C(\vert\psi_{i}\rangle) ,
\end{equation}
where the minimum is taken over all decompositions of $\rho$ into pure states: $\rho= \sum_{i} p_{i} \vert\psi_{i}\rangle\langle\psi_{i}\vert$, where $p_{i}\geq 0$ and $\sum_{i} p_{i}=1$. The state $\rho$ is separable if and only if $C(\rho)=0$. But, unfortunately,  $C(\rho)$ can not be computed, in general. Only for the two-qubit case the problem has been solved, i.e. the minimum in Eq.  (\ref{eq:2}) can be taken, for which we have \cite{15}
\begin{equation}
\label{eq:3}
C(\rho)= \max {\left\lbrace  \Lambda_{1}-\sum_{j>1}^{4}{\Lambda_{j}}, 0\right\rbrace },
\end{equation}
where $\Lambda_{j}$  are the square roots of the eigenvalues of the matrix $R=\rho (\sigma_{y}^{A}\otimes \sigma_{y}^{B})\rho^{\ast} (\sigma_{y}^{A}\otimes\sigma_{y}^{B})$, in decreasing order. $\sigma_{y}$ is the second Pauli matrix and $\rho^{\ast}$ is the complex conjugation
of $\rho$ in the computational basis. 

An important property of the concurrence is that it is an \textit{entanglement monotone} \cite{16}. An entanglement monotone does not increase, on average, under local operations and classical communication (LOCC) \cite{17}. Therefore, if, under LOCC, the initial state $\rho$ transforms to an ensemble of the final states $\left\lbrace q_{k}, \rho_{k}^{\prime}\right\rbrace$, where $\left\lbrace q_{k}\right\rbrace$  is a probability distribution ($q_{k}\geq 0$ and $\sum_{k} q_{k}=1$) and $ \rho_{k}^{\prime}$  are the different possible final states, we have
\begin{equation}
\label{eq:4}
C(\rho)\geq \sum_{k} q_{k} C(\rho_{k}^{\prime}).
\end{equation}
We will use this property of concurrence in Sec. ~\ref{sec:inaccessible}.

\section{Markov states} \label{markov}

 A tripartite state $\rho_{ABE}$ is called a Markov state if  there exists a decomposition of the Hilbert space of the subsystem $B$, $\cH_{B}$, as $\cH_{B}=\bigoplus_{k}\cH_{b^{L}_{k}}\otimes\cH_{b^{R}_{k}}$  such that
\begin{equation}
\label{eq:5}
\rho_{ABE}=\bigoplus_{k}\lambda_{k}\:\rho_{Ab^{L}_{k}}\otimes\rho_{b^{R}_{k}E},
\end{equation}
where $\lbrace \lambda_{k}\rbrace$ is a probability distribution, $\rho_{Ab^{L}_{k}}$ is a state on $\cH_{A}\otimes\cH_{b^{L}_{k}}$ and $\rho_{b^{R}_{k}E}$ is a state on $\cH_{b^{R}_{k}}\otimes\cH_{E}$ \cite{18}. ($\cH_{A}$ and $\cH_{E}$ are the Hilbert spaces of $A$ and $E$, respectively.)

It can be shown that if a tripartite state $\rho_{ABE}$ is a Markov state, then each \textit{localized} dynamics as 
\begin{equation}
\label{eq:6}
\begin{aligned}
\rho_{ABE}^{\prime}=\sum_{j}\left( I_{A}\otimes f_{BE}^{(j)}\right)\,\rho_{ABE}\, \left( I_{A}\otimes f_{BE}^{(j) \dagger}\right),\\
\sum_{j}f_{BE}^{(j) \dagger}f_{BE}^{(j)}=I_{BE},  \qquad \qquad
\end{aligned}
\end{equation}
 reduces to a localized subdynamics as
\begin{equation}
\label{eq:7}
\begin{aligned}
\rho_{AB}^{\prime}=\sum_{i}\left( I_{A}\otimes E_{B}^{(i)}\right)\,\rho_{AB}\, \left( I_{A}\otimes E_{B}^{(i)\dagger}\right), \\
\sum_{i} E_{B}^{(i)\dagger}E_{B}^{(i)}=I_{B},\qquad\qquad
\end{aligned}
\end{equation}
 and vice versa \cite{19, 12, 13}. In Eq. (\ref{eq:6}),
 $f_{BE}^{(j)}$ are linear operators on $BE$ and, in Eq. (\ref{eq:7}),
  $E_{B}^{(i)}$ are linear operators on $B$, 
 $\rho_{AB}=\mathrm{Tr_{E}}(\rho_{ABE})$ is the initial state of  $S=AB$ and $\rho_{AB}^{\prime}=\mathrm{Tr_{E}}(\rho_{ABE}^{\prime})$ is the final state of $S$. In addition, $I_{A}$,  $I_{B}$ and $I_{BE}$
 are the identity operators on $A$, $B$ and $BE$, respectively.
 
 Consider a bipartite system $S=AB$, such that the part $A$ is isolated from the environment $E$ and only the part $B$ interacts with the $E$. So, the evolution of the whole system-environment is as Eq. (\ref{eq:6}). Now, if the initial state of the system-environment $\rho_{ABE}$ is a Markov state as Eq. (\ref{eq:5}), then the reduced dynamics of the system will be localized as Eq. (\ref{eq:7}). Entanglement of the system $S=AB$ does not increase under local operations as Eq. (\ref{eq:7}) \cite{20a}. So, in order to see entanglement exceeding in the system, we must choose the initial state of the system-environment a non-Markovian state. We will use this fact in the next section.

%In the next section, the above property of the Markov states guides us to choose an appropriate initial state for the whole system-environment.

The definition of the Markov states can be generalized to the quadripartite case, too. A quadripartite state $\rho_{AE_{A}BE_{B}}$ is called a Markov state if there exist decompositions of  $\cH_{A}$ and $\cH_{B}$ as $\cH_{A}=\bigoplus_{j}\cH_{a_{j}^{L}}\otimes\cH_{a_{j}^{R}}$ and $\cH_{B}=\bigoplus_{k}\cH_{b_{k}^{L}}\otimes\cH_{b_{k}^{R}}$, such that 

\begin{equation}
\label{eq:8}
\begin{aligned}
\rho_{AE_{A}BE_{B}}
=\bigoplus_{j, k}\lambda_{jk}\,\rho_{a_{j}^{L}E_{A}}\otimes\rho_{a_{j}^{R}b_{k}^{L}}\otimes\rho_{b_{k}^{R}E_{B}}\; ,\\
\end{aligned}
\end{equation}
where  $\lbrace \lambda_{jk}\rbrace$  is a probability distribution, $\rho_{a_{j}^{L}E_{A}}$ is a state on $\cH_{a_{j}^{L}}\otimes\cH_{E_{A}}$, $\rho_{a_{j}^{R}b_{k}^{L}}$ is a state on $ \cH_{a_{j}^{R}}\otimes\cH_{b_{k}^{L}}$ and $\rho_{b_{k}^{R}E_{B}}$ is a state on $\cH_{b_{k}^{R}}\otimes\cH_{E_{B}}$ \cite{12, 13}. ($ \cH_{E_{A}}$ and $ \cH_{E_{B}}$ are the Hilbert spaces of $E_{A}$ and  $E_{B}$, respectively.)

It can be shown that if the initial state $\rho_{AE_{A}BE_{B}}$, of our quadripartite configuration,  is a Markov state as Eq. (\ref{eq:8}), then each localized dynamics as $\mathcal{F}_{AE_{A}}\otimes \mathcal{F}_{BE_{B}}$, on the whole  $AE_{A}BE_{B}$, reduces to a localized subdynamics as $\mathcal{E}_{A}\otimes \mathcal{E}_{B}$, on the $S=AB$,
 where $\mathcal{F}_{AE_{A}}$, $\mathcal{F}_{BE_{B}}$, $\mathcal{E}_{A}$ and $\mathcal{E}_{B}$ are completely positive maps on $AE_{A}$, $BE_{B}$, $A$ and $B$, respectively \cite{12, 13}. (A completely positive map, on a state $\rho$ , is a map which can be written as 
$\sum_{i} K_{i}\rho K_{i}^{\dagger}$, where $K_{i}$ are linear operators such that 
$\sum_{i} K_{i}^{\dagger} K_{i}=I$  ($I$ is the identity operator) \cite{20}.)

Note that if the reduced dynamics of  $S=AB$ can be written as $\mathcal{E}_{A}\otimes \mathcal{E}_{B}$, then the entanglement of $S$ does not increase under such localized evolution.
 Therefore,  if, for a localized dynamics  $\mathcal{F}_{AE_{A}}\otimes \mathcal{F}_{BE_{B}}$ on the whole $AE_{A}BE_{B}$, the entanglement of  $S$ increases, then we conclude that the initial state of $AE_{A}BE_{B}$ is not a Markov state as Eq. (\ref{eq:8}).

For example, for the case studied in Ref. \cite{2b},
% let's denote the two separated reservoirs as $A$ and $B$, and the two separated cavities as $E_{A}$ and $E_{B}$. In Ref. \cite{6}, the initial state of the whole  $AE_{A}BE_{B}$ has been chosen as
%\begin{equation}
%\label{eq:8a}
%\vert\Phi_{0}\rangle= \vert 0_{A}\rangle\otimes\vert 0_{B}\rangle\otimes \vert\Psi_{E_{A}E_{B}}\rangle,
%\end{equation}
%where $\vert\Psi_{E_{A}E_{B}}\rangle$ is an entangled state in $E_{A}E_{B}$ and $\vert 0_{A}\rangle$ and $\vert 0_{B}\rangle$ are some fixed states in $A$ and $B$, respectively.
  the initial state $\vert\Phi_{0}\rangle$, in Eq.  (\ref{eq:1a}), is not a Markov state. From Eq. (\ref{eq:8}), we know that, for a Markov state $\rho_{AE_{A}BE_{B}}$,   $\rho_{E_{A}E_{B}}=\mathrm{Tr_{AB}}(\rho_{AE_{A}BE_{B}})$ is a separable state, but $\vert\Psi_{E_{A}E_{B}}\rangle$ is entangled. Therefore, though the whole dynamics of  $AE_{A}BE_{B}$ is localized as $\mathcal{F}_{AE_{A}}\otimes \mathcal{F}_{BE_{B}}$, the reduced dynamics of $S=AB$, can be non-localized and lead to exceeding the entanglement of $S$.  This is in agreement with the result of Ref. \cite{2b}.
%Note that the same argument can be given for the case studied in  Ref. \cite{2b}, with the initial
% system-environment  state given in Eq.  (\ref{eq:1a}).

%The following point is also worth noting. Though the whole dynamics of the system-environment is localized, but the initial state of the environment in Eq.  (\ref{eq:8a}), i.e. $\vert\Psi_{E_{A}E_{B}}\rangle$, is entangled. In other words, the entangled state $\vert\Psi_{E_{A}E_{B}}\rangle$ is shared between the two separate parts $AE_{A}$ and $BE_{B}$. So, we expect that this initially shared entanglement can lead to entanglement increasing in the system. From this point of view, entanglement exceeding in the system $S=AB$ is not unexpected. (The interesting result
% of Ref. \cite{6} is that this entanglement exceeding in the two-reservoir system $S$ can occur as a sudden birth; i.e. the entanglement of the $S=AB$, which is initially zero, remains zero for a time interval and then suddenly appears at some $t_{1}>0$.) 
%
%In the next section, we will study an example, in which, though there is no (initial) entanglement in the environment, the entanglement of the system exceeds its initial value.

It is also worth noting that the non-Markovianity of the initial state of the  system-environment, though a  necessary condition,  is not  sufficient for entanglement exceeding, in general. For example, for the case studied in Ref. \cite{11a}, which we will discuss in detail in the next section, 
 the first and the second initial states of the system-environment, considered there, are not  Markov states.
But, as has been shown in  Ref. \cite{11a}, their second initial state of the  system-environment does not lead to  entanglement exceeding.

\section{ model and  results} \label{sec:model}

%The model and the results 

We consider a bipartite system $S=AB$, including two separated spin-$1/2$ particles.
In addition, for simplicity, we assume that the spin (qubit) $A$ is isolated from the environment and only the spin (qubit) $B$ interacts with its local environment $E$, which includes $N$ spin-$1/2$ particles, through the (interaction) Hamiltonian 
\begin{equation}
\label{eq:9}
H=\sum_{i=1}^{N}g(\sigma_{+}^{B}I_{-}^{(i)}+\sigma_{-}^{B}I_{+}^{(i)}),
\end{equation}
where $\sigma_{\pm}^{B}$ and $I_{\pm}^{(i)}$ are the raising and lowering operators for the spin $B$ and the $i$-th environmental spin, respectively, and $g$ denotes the coupling strength between the spin $B$ and each spin in the environment.
Physically, the above Hamiltonian can describe the hyperfine interaction between an electron spin, confined in a quantum dot, with spins of its surrounding nuclei \cite{11a, 20b}.  

Therefore, the whole dynamics of the system-environment is given by the localized dynamics $U_{SE}(t)=I_{A}\otimes U_{BE}(t)$, where $U_{BE}(t)=e^{-iHt/\hbar}$. So, according to the previous section, to observe entanglement exceeding, we must choose the initial state of the whole system-environment a state which is not a Markov state as Eq. (\ref{eq:5}).

In Ref. \cite{11a}, the entanglement dynamics of the above system $S=AB$, for three different initial states of the system-environment, has been studied. In the first case, they chose the initial state of the system-environment as
\begin{equation}
\label{eq:9a}
\begin{aligned}
\vert\omega_{0}\rangle= x\vert 0_{A}\rangle\vert 0_{B}\rangle \vert\overline{1}_{E} \rangle
+ y\vert 0_{A}\rangle\vert 1_{B}\rangle \vert\overline{0}_{E} \rangle
+ z\vert 1_{A}\rangle\vert 0_{B}\rangle \vert\overline{0}_{E} \rangle,
\end{aligned}
\end{equation}
where $x$, $y$ and $z$ are real nonzero coefficients, such that $x^{2}+y^{2}+z^{2}=1$.  $\vert 0\rangle$ and $\vert 1\rangle$ denote the spin down and the spin up states of the particle, respectively. In addition,  $\vert \overline{0}_{E}\rangle=\vert 00\ldots 0\rangle$  (with $N$ spin down) and $\vert \overline{1}_{E}\rangle=\frac{1}{\sqrt{N}}\sum_{i=1}^{N}I_{+}^{(i)}\vert \overline{0}_{E}\rangle$ are  the ground and the first excitation states of the environment, respectively.
$\vert\omega_{0}\rangle$ is a $W$-class genuine tripartite entangled state \cite{20c}. It can be shown simply that 
$\rho_{AE}= \mathrm{Tr_{B}}(\vert\omega_{0}\rangle\langle\omega_{0}\vert)$ is entangled.  Equation (\ref{eq:5})  results in, for a Markov state $\rho_{ABE}$, the reduced state $\rho_{AE}= \mathrm{Tr_{B}}(\rho_{ABE})$ being separable. Therefore, $\vert\omega_{0}\rangle$  is not a Markov state.
So, entanglement exceeding  can occur in this case. This is in agreement with the results of Ref. \cite{11a}.

The third initial state of the system-environment, considered in Ref. \cite{11a}, is a factorized state as $\rho_{ABE}=\rho_{AB}\otimes\rho_{E}$. So, it is a Markov state, which is due to the case that, in Eq. (\ref{eq:5}),  $\cH_{B}=\cH_{b^{L}}\otimes\cH_{b^{R}}$ and $\cH_{b^{R}}$ is a trivial one-dimensional Hilbert space. Therefore, entanglement of the system $S=AB$ never exceeds its initial value. This is also in agreement with the results of Ref. \cite{11a}.

Note that, for the initial state $\vert\omega_{0}\rangle$ in Eq. (\ref{eq:9a}), obviously, there is no entanglement in the environment, since the environment is simply one-partite. So, the first case studied in Ref. \cite{11a}, in fact, shows that 
 the entanglement of the system can exceed its initial value, even if the  dynamics of the whole system-environment is localized and there is no initial entanglement in the environment.
 
 However, the initial state of the whole system-environment $\vert\omega_{0}\rangle$, in Eq. (\ref{eq:9a}), is entangled. In other words, initially, there is non-classical correlation between the system and the environment. In the following, we choose the initial state for the  system-environment such that, at the initial moment, there is only classical correlation between the system and the environment and there is no entanglement in the environment, but, interestingly, entanglement exceeding in the system occurs for it.

Instead of the second initial state of the system-environment, considered in Ref. \cite{11a}, we choose the following state as our initial state of the system-environment:
\begin{equation}
\label{eq:11}
\begin{aligned}
\rho_{SE}=\rho_{ABE}=p\vert\psi_{AB}^{(1)}\rangle\langle\psi_{AB}^{(1)}\vert\otimes\vert \overline{1}_{E}\rangle\langle \overline{1}_{E}\vert \qquad\qquad  \\
+(1-p)\vert\psi_{AB}^{(2)}\rangle\langle\psi_{AB}^{(2)}\vert\otimes\vert \overline{0}_{E}\rangle\langle \overline{0}_{E}\vert,
\end{aligned}
\end{equation}
where $0\leq p\leq1$ and
\begin{equation}
\label{eq:10}
\begin{aligned}
\vert\psi_{AB}^{(1)}\rangle= \cos\alpha\vert 1_{A}\rangle\vert 0_{B}\rangle + \sin\alpha\vert 0_{A}\rangle\vert 1_{B}\rangle,  \\
\vert\psi_{AB}^{(2)}\rangle= \cos\beta\vert 1_{A}\rangle\vert 1_{B}\rangle + \sin\beta\vert 0_{A}\rangle\vert 0_{B}\rangle,
\end{aligned}
\end{equation}
where $0\leq \alpha, \beta \leq 2 \pi$.
 Note that $\langle\psi_{AB}^{(2)}\vert\psi_{AB}^{(1)}\rangle=0$. In addition,  
 $\langle \overline{0}_{E}\vert \overline{1}_{E}\rangle=0$. So, the system $S$ and the environment $E$ are only classically correlated; i.e. the \textit{quantum discord} of the (bipartite) state $\rho_{SE}$,    in Eq. (\ref{eq:11}), is zero \cite{21a}.

 According to Eq. (\ref{eq:9}), we see that the time evolution operator $U_{SE}$ preserves the excitations of the whole system-environment. In addition, the initial state $\rho_{SE}$, in Eq. (\ref{eq:11}), is a state on the subspace spanned by  $\lbrace\vert 0_{A}0_{B}\overline{0}_{E}\rangle, \vert 1_{A}1_{B}\overline{0}_{E}\rangle, \vert 0_{A}1_{B}\overline{1}_{E}\rangle, \vert 1_{A}0_{B}\overline{1}_{E}\rangle, \vert 0_{A}0_{B}\overline{2}_{E}\rangle\rbrace$,  where  $\vert \overline{2}_{E}\rangle =\sqrt{\frac{2}{N(N-1)}}\sum_{i,j=1;i<j}^{N}I_{+}^{(i)}I_{+}^{(j)}\vert \overline{0}_{E}\rangle$.  The state $\vert 0_{A}0_{B}\overline{0}_{E}\rangle$ spans the one-dimensional subspace of
 zero excitation. So it is invariant during the evolution. 
 The states $\vert 1_{A}1_{B}\overline{0}_{E}\rangle$,  $\vert 0_{A}1_{B}\overline{1}_{E}\rangle$,  $\vert 1_{A}0_{B}\overline{1}_{E}\rangle$ and $\vert 0_{A}0_{B}\overline{2}_{E}\rangle$ span the subspace of two excitations.
 The restriction  of $U_{SE}$, on this four-dimensional subspace, has been given in Ref. \cite{11a}.
 Therefore, we can simply obtain the reduced state of the system, at the time $t$, as $\rho_{AB}(t)=\mathrm{Tr_{E}}(U_{SE}(t)\rho_{SE}U_{SE}^{\dagger}(t))$.
 
  Finally, using Eq. (\ref{eq:3}), we obtain the concurrence of our two-qubit system as 
\begin{equation}
\label{eq:11a}
\begin{aligned}
C(\rho_{AB}(t))=2 \mathrm{max}\lbrace 0,C_{1}(t),C_{2}(t)\rbrace, \qquad \\ 
%%\qquad
C_{1}(t)=\vert e \cos(\Omega_{1}t)\cos(\Omega t)\vert  \qquad\quad\quad\quad    \\ 
-\sqrt{(b \cos^{2}(\Omega t)+f \sin^{2}(\Omega t))(a+d \sin^{2}(\Omega_{1}t))}, \\ 
C_{2}(t)=\vert c \cos(\Omega t)\vert \qquad\quad \qquad\quad \qquad\quad \\
-\sqrt{(b \sin^{2}(\Omega t)+f \cos^{2}(\Omega t))d \cos^{2}(\Omega_{1}t)},
\end{aligned}
\end{equation} 
where $a=(1-p)\sin^{2}\beta$, $b=(1-p)\cos^{2}\beta$, $c=0.5(1-p)\sin 2\beta$, $d=p\sin^{2}\alpha$, $e=0.5p\sin 2\alpha$, $f=p\cos^{2}\alpha$, $\Omega=g\sqrt{N}$ and $\Omega_{1}=g\sqrt{2N-2}$.

\begin{figure}
\begin{center}
\includegraphics[width=8.5cm]{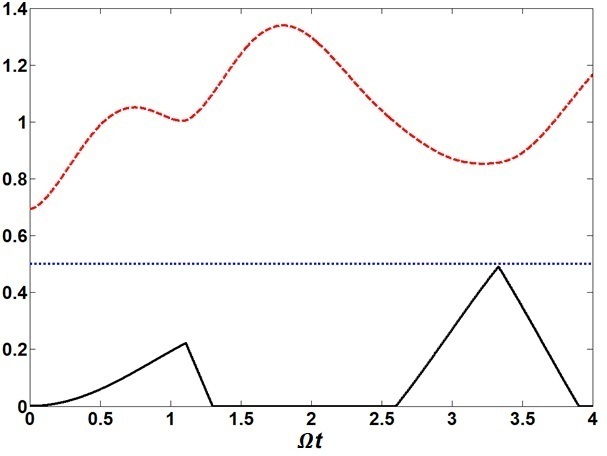}
\end{center}
\caption{Concurrence (black solid line), mutual information (red dashed line) and $C(\rho_{A;BE})$, given in Eq. (\ref{eq:21}), (blue dotted line), as the functions of the scaled time $\Omega t$, for $p=0.5$ and $\alpha=\beta=\pi/4$.}
\label{Fig1}
\end{figure}

In Fig.~\ref{Fig1}, the concurrence of the system (the black solid line) is plotted as the function of the scaled time $\Omega t$, for $p=0.5$ and $\alpha=\beta=\pi/4$. As we see, the concurrence starts to exceed just from the initial moment.

In example 2 of Ref. \cite{12}, we have shown that $\rho_{ABE}$ in Eq. (\ref{eq:11}), for $\alpha\neq n\frac{\pi}{2}$ or $\beta\neq n\frac{\pi}{2}$ ($n=0, ..., 4$),  is not a Markov state, i.e. it can not be written as Eq. (\ref{eq:5}). Therefore, though the whole dynamics of the system-environment is localized, the reduced dynamics of the system $S=AB$, can be non-localized and so lead to exceeding the entanglement of the system, as illustrated in Fig ~\ref{Fig1}. 

Note that, since the whole dynamics of the system-environment is as $I_{A}\otimes U_{BE}(t)$, the reduced state of the qubit $A$ remains unchanged during the evolution. But, e.g., during the time interval $\left( \Omega t_{1}=0, \Omega t_{2}=1.111\right]$ for which the concurrence of the system increases monotonically, the reduced dynamics of the system $S=AB$ is not equivalent  to any localized map as $id_{A}\otimes\mathcal{E}_{B}$, where $id_{A}$ is the identity map on $A$ and $\mathcal{E}_{B}$ is a completely positive map on $B$.
In fact, it is not equivalent to any local operation and, even,  any LOCC map.

But, interestingly, the reduced dynamics of the system can be represented by a completely positive map, using the result of Ref. \cite{21}. Let's define
\begin{equation}
\label{eq:12}
\begin{aligned}
\vert\psi_{AB}^{(3)}\rangle= \sin\alpha\vert 1_{A}\rangle\vert 0_{B}\rangle- \cos\alpha\vert 0_{A}\rangle\vert 1_{B}\rangle,  \\
\vert\psi_{AB}^{(4)}\rangle= \sin\beta\vert 1_{A}\rangle\vert 1_{B}\rangle - \cos\beta\vert 0_{A}\rangle\vert 0_{B}\rangle.
\end{aligned}
\end{equation}
From Eqs. (\ref{eq:10}) and (\ref{eq:12}), we see that $\lbrace \vert\psi_{AB}^{(i)}\rangle\rbrace$ is an orthonormal basis for our two-qubit system $S=AB$. In addition, we define $\vert \mu_{E}^{(1)}\rangle=\vert\overline{1}_{E}\rangle$ and  $\vert \mu_{E}^{(2)}\rangle=\vert \overline{0}_{E}\rangle$ and we choose $\vert \mu_{E}^{(3)}\rangle$ and $\vert \mu_{E}^{(4)}\rangle$, arbitrarily. Therefore, $\rho_{SE}$ in Eq. (\ref{eq:11}) can be rewritten as
\begin{equation}
\label{eq:13}
\begin{aligned}
\rho_{SE}=\rho_{ABE}=\sum_{i=1}^{4} p_{i}\vert\psi_{AB}^{(i)}\rangle\langle\psi_{AB}^{(i)}\vert\otimes\vert \mu_{E}^{(i)}\rangle\langle \mu_{E}^{(i)}\vert,
\end{aligned}
\end{equation}
with $p_{1}=p$, $p_{2}=1-p$ and $p_{3}=p_{4}=0$. 

Note that, since $ \vert\psi_{AB}^{(i)}\rangle$  are orthonormal,
 the system $S$ and the environment $E$ are only classically correlated. In other words, the quantum discord (between the system $S=AB$ and the environment $E$) for any bipartite state $\rho_{SE}$,  which can be written as Eq. (\ref{eq:13}), is zero \cite{21a}.

Now, it has been shown in Ref. \cite{21} that if the initial state of the whole system-environment is as Eq. (\ref{eq:13}), with arbitrary probability distribution $\lbrace p_{i}\rbrace$, but fixed $\vert\psi_{AB}^{(i)}\rangle$ and $\vert \mu_{E}^{(i)}\rangle$, then, for any arbitrary completely positive dynamics for the whole  system-environment, the reduced dynamics of the system is given by a completely positive map. For example, in our case, for which the whole dynamics of the system-environment is given by $I_{A}\otimes U_{BE}$, the reduced dynamics of the system $S=AB$ is given by the completely positive map
\begin{equation}
\label{eq:14}
\begin{aligned}
\rho_{AB}^{\prime}=\sum_{i,k} E_{ik}\,\rho_{AB} E_{ik}^{\dagger}, \\
\sum_{i,k} E_{ik}^{\dagger}E_{ik}=I_{AB},
\end{aligned}
\end{equation}
where $\rho_{AB}=\mathrm{Tr_{E}}(\rho_{ABE})$ is the initial state of the system  and $\rho_{AB}^{\prime}=\mathrm{Tr_{E}}(\rho_{ABE}^{\prime})$ is the final state of the system ($\rho_{ABE}^{\prime}=I_{A}\otimes U_{BE}\rho_{ABE}I_{A}\otimes U_{BE}^{\dagger}$). In addition, $E_{ik}=D_{ik}\Pi_{i}$ are linear operators on $S=AB$, where $D_{ik}=I_{A}\otimes \langle k_{E}\vert U_{BE}\vert\mu_{E}^{(i)}\rangle$, $\Pi_{i}=\vert\psi_{AB}^{(i)}\rangle\langle\psi_{AB}^{(i)}\vert$ and $\lbrace\vert k_{E}\rangle\rbrace$ is an orthonormal basis for the environment $E$. Note that, though $D_{ik}$ is localized, but, because of the factor $\Pi_{i}$, $E_{ik}$ is not so.

%In summary, we have shown that, under localized dynamics of the whole system-environment, the entanglement of the system can exceed its initial value, if the initial state of the system-environment is as Eq. (\ref{eq:11}), which is not a Markov state. In addition, the reduced dynamics of the system is given by the completely positive map in Eq. (\ref{eq:14}).
%
%As far as we know, this is the first time that entanglement exceeding in the system, under local interaction with the environment, is shown for a case which  possesses the three following  features, simultaneously:
%
%1- The (initial) entanglement of the environment is zero.
%
%2- The  initial state of the  system-environment contains only classical correlation between the system and the environment.
%
%3- The reduced dynamics of the system is completely positive.

%In the next section, we will give an explanation of this result, introducing the concept of \textit{inaccessible entanglement}.

Let's end this section with an additional remark.
The \textit{mutual information} of a bipartite state $\rho_{AB}$ is defined as $I(A:B)_{\rho}=S(\rho_A)+S(\rho_B)-S(\rho_{AB})$, where $\rho_{A}=\mathrm{Tr_{B}}(\rho_{AB})$ and $\rho_{B}=\mathrm{Tr_{A}}(\rho_{AB})$ are the reduced states and $S(\rho)$ is the von Neumann entropy of the state $\rho$: $S(\rho)=- \mathrm{Tr}(\rho\log\rho)$ \cite{20}.
% The mutual information $I(A:B)_{\rho^{\prime}}$ of the final state $\rho_{AB}^{\prime}$ is defined similarly.
 Now, in the theorem 11.15 of Ref. \cite{20}, it has been shown that if  $\rho_{AB}^{\prime}=id_{A}\otimes\mathcal{E}_{B}(\rho_{AB})$, where $id_{A}$ is the identity map on $A$ and $\mathcal{E}_{B}$ is a completely positive map on $B$, then
\begin{equation}
\label{eq:15}
I(A:B)_{\rho}\geq I(A:B)_{\rho^{\prime}}.
\end{equation} 
Now, from Fig. ~\ref{Fig1}, we see that, e.g., for all  $\Omega t^{\prime}_{1}\in\left(\Omega t_{1}=2.603, \Omega t_{2}=3.333\right]$ for which the concurrence of the system increases monotonically and so $\rho_{AB}(t^{\prime}_{1})\neq id_{A}\otimes\mathcal{E}_{B}(\rho_{AB}(t_{1}))$, the mutual information  decreases. Therefore, the reverse of the above theorem is not  valid, in general: When the mutual information $I(A:B)$ decreases, we can not conclude that the dynamics of the system is equivalent to a localized dynamics as $id_{A}\otimes\mathcal{E}_{B}$.
(In Fig. ~\ref{Fig1}, we have plotted the mutual information using $\mathrm{log_{10}}$  instead of $\mathrm{log_{2}}$. This is equivalent to multiplication by a (less than 1) positive constant which makes the mutual information of a similar order as  the concurrence and improves the comparison between them.)

\section{Inaccessible entanglement}\label{sec:inaccessible}
%Inaccessible entanglement

Consider the case that the two parts $A$ and $B$ of our bipartite system, are separated from each other and each part interacts with its own local environment. We denote the local environment of $A$ as $E_{A}$, the local environment of $B$ as $E_{B}$ and the whole state of the system-environments as $\rho_{AE_{A};BE_{B}}$. Therefore, our quadripartite configuration consists of two separated parts $AE_{A}$ and $BE_{B}$. Let's define \textit{inaccessible entanglement} as 
\begin{equation}
\label{eq:16}
\mathcal{M}_{I}=\mathcal{M}(\rho_{AE_{A};BE_{B}})-\mathcal{M}(\rho_{A;B}),
\end{equation} 
where $\mathcal{M}$ is an appropriate entanglement measure (monotone) and $\rho_{A;B}=\mathrm{Tr_{E_{A}E_{B}}}(\rho_{AE_{A};BE_{B}})$. Note that, since $\mathcal{M}$ is an entanglement monotone defined for bipartite systems, $\mathcal{M}(\rho_{AE_{A};BE_{B}})$ is calculated according to the  bipartition $({AE_{A}; BE_{B}})$. In addition, since the partial traces over $E_{A}$ and $E_{B}$ are  local operations \cite{17,20}, $\mathcal{M}(\rho_{AE_{A};BE_{B}})\geq\mathcal{M}(\rho_{A;B})$ and so we always have $\mathcal{M}_{I}\geq 0$.

Assuming that we have access only to the system and not to the environments, the meaning of the inaccessible entanglement is clear: It measures the amount of entanglement which is present between the two separated parts $AE_{A}$ and $BE_{B}$, but is inaccessible for us. If, at the initial moment,  we have $\mathcal{M}_{I}> 0$, it means that there is a supply of entanglement, in the whole system-environment, which is inaccessible, initially.
But, during the interaction of the system and the environments, even if this interaction is localized, (a part of) this supply can transfer to the system and lead to exceeding the entanglement of the system,   rather than its initial value.

In this paper, we use the concurrence as the entanglement measure (monotone). So, we rewrite Eq. (\ref{eq:16}) as
\begin{equation}
\label{eq:17}
C_{I}(t)=C\left(\rho_{AE_{A};BE_{B}}(t)\right)-C\left(\rho_{A;B}(t)\right).
\end{equation} 

Let's consider the case studied in the previous section.
First, note that, from Eq. (\ref{eq:1}), we have
\begin{equation}
\label{eq:18}
\begin{aligned}
C(\vert\psi_{AB}^{(1)}\rangle\otimes\vert \overline{1}_{E}\rangle)=C(\vert\psi_{AB}^{(1)}\rangle), \\
C(\vert\psi_{AB}^{(2)}\rangle\otimes\vert \overline{0}_{E}\rangle)=C(\vert\psi_{AB}^{(2)}\rangle),
\end{aligned}
\end{equation}
where the concurrence is calculated according to the bi-partition $(A; BE)$. So, for the initial state of the system-environment, given in Eq. (\ref{eq:11}), according to Eq. (\ref{eq:2}), we have
\begin{equation}
\label{eq:19}
C(\rho_{A;BE})\leq pC(\vert\psi_{AB}^{(1)}\rangle)+(1-p)C(\vert\psi_{AB}^{(2)}\rangle).
\end{equation} 
On the other hand, performing the local projective measurement, given by $\lbrace I_{AB}\otimes\vert \overline{0}_{E}\rangle\langle \overline{0}_{E}\vert, I_{AB}\otimes\vert \overline{1}_{E}\rangle\langle \overline{1}_{E}\vert , I_{AB}\otimes(I_{E}-\vert \overline{0}_{E}\rangle\langle \overline{0}_{E}\vert-\vert \overline{1}_{E}\rangle\langle \overline{1}_{E}\vert)\rbrace$, the initial state, given in Eq. (\ref{eq:11}), transforms to the ensemble $\lbrace (p, \vert\psi_{AB}^{(1)}\rangle\otimes\vert \overline{1}_{E}\rangle), (1-p, \vert\psi_{AB}^{(2)}\rangle\otimes\vert \overline{0}_{E}\rangle)\rbrace$. Therefore, from Eqs. (\ref{eq:4}) and (\ref{eq:18}), we have
\begin{equation}
\label{eq:20}
C(\rho_{A;BE})\geq pC(\vert\psi_{AB}^{(1)}\rangle)+(1-p)C(\vert\psi_{AB}^{(2)}\rangle).
\end{equation}
So, combining  Eqs. (\ref{eq:19}) and (\ref{eq:20})  gives us
\begin{equation}
\label{eq:21}
C(\rho_{A;BE})= pC(\vert\psi_{AB}^{(1)}\rangle)+(1-p)C(\vert\psi_{AB}^{(2)}\rangle).
\end{equation}
In addition, since the whole dynamics of the system-environment is given by the local unitary transformation $I_{A}\otimes U_{BE}(t)$, we have $C(\rho_{A;BE}(t))=C(\rho_{A;BE}(0))$. Therefore, Eq. (\ref{eq:17}) can be rewritten as
\begin{equation}
\label{eq:22}
C_{I}(t)=pC(\vert\psi_{AB}^{(1)}\rangle)+(1-p)C(\vert\psi_{AB}^{(2)}\rangle)-C\left(\rho_{A;B}(t)\right).
\end{equation} 
In Fig. ~\ref{Fig1},   $C(\rho_{A;BE}(t))=C(\rho_{A;BE}(0))=pC(\vert\psi_{AB}^{(1)}\rangle)+(1-p)C(\vert\psi_{AB}^{(2)}\rangle)$ is plotted, as the blue dotted line. So, $C_{I}(t)$ is given by the difference between this blue dotted line and the black solid curve, which gives the $ C\left(\rho_{A;B}(t)\right)$. Since 
$C_{I}(0)>0$, there is an initial supply of entanglement, in the whole system-environment, which is initially inaccessible for the system. As we see from Fig. ~\ref{Fig1}, during the time evolution, (a part of) this supply can transfer to the system and lead to  the exceeding   $C\left(\rho_{A;B}(t)\right)$ than 
 $C\left(\rho_{A;B}(0)\right)$, for some times $t$.
 
 Note that, as mentioned in the previous section, since the environment is one-partite, the (initial) supply of the entanglement in the environment is zero. But, interestingly, the initial supply of the entanglement in the \textit{ whole system-environment}, which is initially inaccessible for the system, i.e. $C_{I}(0)$, is greater than zero and leads to entanglement exceeding, in the system.

%It seems that  this interesting result is due this important feature of  composite systems  that the complete description of the subsystems does not give us  the complete description of the whole system. In general, it is due to both classical and quantum correlations between the subsystems. But, interestingly, in our case, only classical correlation between the system and the environment leads to the non-completenesity of the subsystems  description for a feature which is considered non-classical, i.e. the entanglement.

It is also worth noting that the previously introduced concept of \textit{hidden entanglement}, in Ref. \cite{22}, is, in fact, a special case of the inaccessible entanglement, introduced here. This can be shown simply, using a result of Ref. \cite{23}.
Hidden entanglement has been introduced to explain the entanglement revival in a system which interacts with a classical environment.
 For example, consider a bipartite quantum system $S=AB$, such that the part $A$ is isolated from the environment and only the part $B$ interacts with a random classical  field. The effect of the random classical  field on $B$ can be modeled as acting random unitary operators $U^{(j)}_{B}$ on $B$, each with the probability $p_{j}$ \cite{23}. Therefore, the whole dynamics of the system can be written as \cite{23}:

\begin{equation}
\label{eq:23}
\begin{aligned}
\rho_{AB}(t)=\sum_{j}p_{j}\, \left( I_{A}\otimes U^{(j)}_{B}(t)\right)\,\rho_{AB}(0)\, \left( I_{A}\otimes U^{(j)\dagger}_{B}(t)\right),
%\quad \\ 
\end{aligned}
\end{equation}
where  $\rho_{AB}(0)$ is the initial state of the system and $\rho_{AB}(t)$ is the state of the system at time $t$.

 We can model the whole system-environment evolution as the following \cite{23}. We get the initial state of the system-environment as
 \begin{equation}
\label{eq:24}
\begin{aligned}
\rho_{SE}(0)=\rho_{AB}(0)\otimes\sum_{j}p_{j}\vert j_{E}\rangle\langle j_{E}\vert ,
\end{aligned}
\end{equation} 
where $\lbrace\vert j_{E}\rangle\rbrace$ is an orthonormal basis for the environment $E$. In addition, the system-environment undergoes the evolution given by the unitary operator
 \begin{equation}
\label{eq:25}
\begin{aligned}
U_{SE}(t)=\sum_{j} I_{A}\otimes U^{(j)}_{B}(t) \otimes\vert j_{E}\rangle\langle j_{E}\vert .
\end{aligned}
\end{equation}
From Eqs. (\ref{eq:24}) and (\ref{eq:25}), it can be shown simply that the reduced dynamics of the system $S=AB$ is given by Eq. (\ref{eq:23}). In addition, the reduced state of the environment remains unchanged during the evolution, as expected. We have $\rho_{E}(t)=\sum_{j}p_{j}\vert j_{E}\rangle\langle j_{E}\vert=\rho_{E}(0)$, which is a classical state, i.e., it contains no superposition of the basis states $\vert j_{E}\rangle$.

First, note that, since the initial state $\rho_{A;BE}(0)$ in Eq. (\ref{eq:24}) is factorized, it can be shown simply that $C(\rho_{A;BE}(0))=C(\rho_{A;B}(0))$.
 In addition, since the dynamics of the whole system-environment is given by the local unitary transformation, given in Eq. (\ref{eq:25}), we have
  \begin{equation}
\label{eq:26}
\begin{aligned}
C(\rho_{A;BE}(t))=C(\rho_{A;BE}(0))=C(\rho_{A;B}(0)).
\end{aligned}
\end{equation}
  
On the other hand, if we define
\begin{equation}
\label{eq:27}
\begin{aligned}
\rho_{A;B}^{(j)}(t)
= \left( I_{A}\otimes U^{(j)}_{B}(t)\right)\,\rho_{AB}(0)\, \left( I_{A}\otimes U^{(j)\dagger}_{B}(t)\right),
\end{aligned}
\end{equation}
then, since under local unitary transformations entanglement does not change, we have $C(\rho_{A;B}^{(j)}(t))= C(\rho_{A;B}(0))$. Therefore, we can rewrite Eq. (\ref{eq:26}) as
\begin{equation}
\label{eq:28}
\begin{aligned}
C(\rho_{A;BE}(t))=C(\rho_{A;B}(0))=\sum_{j}p_{j}C(\rho_{A;B}^{(j)}(t)).
\end{aligned}
\end{equation}

Finally, similar to Eqs. (\ref{eq:16}) and (\ref{eq:17}), we have
\begin{equation}
\label{eq:29}
\begin{aligned}
C_{I}(t)=C\left(\rho_{A;BE}(t)\right)-C\left(\rho_{A;B}(t)\right) \\
=\sum_{j}p_{j}C(\rho_{A;B}^{(j)}(t))-C\left(\rho_{A;B}(t)\right),
\end{aligned}
\end{equation} 
which coincides with the definition of the hidden entanglement, given in Ref. \cite{22}. (Note that Eqs. (\ref{eq:16}) and (\ref{eq:17}) are written for the quadripartite configuration, but Eq. (\ref{eq:29}) is for the tripartite configuration.)

In this case, since $C_{I}(0)=0$, there is no initial supply of entanglement in the system-environment which is inaccessible for the system. So, the entanglement of the system can not exceed its initial value. In this case, only the entanglement revival can occur; i.e. for some times $t>0$ the entanglement of the system can reach its initial value, but can not exceed it. This is in agreement with the results of Refs. \cite{22, 23}.

It is also worth noting that, in fact, there are two minor differences between the inaccessible entanglement in Eq. (\ref{eq:29}) and the definition of hidden entanglement, given in Ref. \cite{22}. First, there, \textit{entanglement of formation} \cite{1, 20a} is used as the entanglement measure, instead of concurrence which we used here. Entanglement of formation is also an entanglement monotone \cite{16,17} and so a similar line of reasoning, similar to that given from Eqs.  (\ref{eq:26})-(\ref{eq:29}), can be given for it, too. Second, there, the definition of the hidden entanglement is restricted to the case that (the initial state of the system is pure and so) the final ensemble is an ensemble of pure states. Here, it is generalized  to include the case that (the initial state of the system is mixed and so) the final ensemble is an ensemble of mixed states as $\lbrace p_{j}, \rho_{A;B}^{(j)}(t)\rbrace$.

\section{Conclusion}\label{sec:summary}
%Summary and discussion
%Conclusion

In this paper, we have considered the case that the system $S=AB$ is  bipartite and the part $A$ is isolated from the environment and  only the part $B$ of the system interacts with its local environment $E$.
We have focused on the phenomenon of exceeding the entanglement, rather than its initial value, in such system.
 
 First, using the results of Refs. \cite{12, 13}, we have emphasized that the phenomenon of entanglement exceeding in the system,  under local interactions with the environment, can occur only when the initial state of the whole system-environment is not a Markov state as Eqs. (\ref{eq:5}) or (\ref{eq:8}).

Second, we have shown that this phenomenon  can occur even if  we have
 the three following   features, simultaneously:

(1) The (initial) entanglement of the environment is zero.

(2) The  initial state of the system-environment contains only classical correlation between the system and the environment.

(3) The reduced dynamics of the system is completely positive.

%As far as we know, this is the first time that entanglement exceeding in an open quantum system, possessing such  features simultaneously, is shown.

Finding this interesting case is not only due the interaction model, considered in Sec.   ~\ref{sec:model} as Eq. (\ref{eq:9}), but also, due choosing the initial state of the system-environment $\rho_{SE}(0)$, appropriately, as Eq. (\ref{eq:11}).
  In  Fig.  ~\ref{Fig1}, if we change the initial moment from $t_{0}=0$ to another  $t_{0}>0$, the dynamics of the entanglement, for $t\geq t_{0}$, may not possess the above three features, simultaneously. However, this does not change the interesting fact that when we choose the initial moment as $t_{0}=0$ [when we choose the initial state as Eq. (\ref{eq:11})], the dynamics of the entanglement, for $t\geq 0$, possesses all the  three above features, simultaneously.

And third, we have given an explanation of  entanglement exceeding, introducing the concept of inaccessible entanglement $C_{I}$. If, at the initial moment, $C_{I}>0$, this means that there is an initial supply of entanglement in the\textit{ whole system-environment}, which is initially inaccessible for the system.  Transfer of (a part of) this supply to the system, during the interaction of the system and the environment, leads to exceeding the entanglement of the system, rather than its initial value.

The applicability of the inaccessible entanglement $C_{I}$ is not restricted to the  case studied in Sec.  ~\ref{sec:model}. This concept can be used to explain entanglement exceeding and entanglement revival, in any open quantum system, interacting with the environment locally. For example, we have shown that the previously introduced concept of hidden entanglement in Ref. \cite{22}, which was introduced to explain entanglement revival when the system is interacting locally with a classical environment, is a special case of the inaccessible entanglement, introduced in this paper.

%\section*{Acknowledgments}
%
%We would like to thank the anonymous referees for their helpful comments.

%%%%%%%%%%%%%%%%%%%%%%%%%%


\begin{thebibliography}{000}

 \bibitem{1} L. Aolita, F. de Melo and L. Davidovich,  Open-system dynamics of entanglement: a key issues review, \href{http://iopscience.iop.org/article/10.1088/0034-4885/78/4/042001/meta} {Rep. Prog. Phys. {\bf 78}, 042001 (2015)}. 
 
  \bibitem{2} T. Yu and J. H. Eberly, Sudden death of entanglement, \href{http://science.sciencemag.org/content/323/5914/598} {Science \textbf{323}, 598 (2009)}.
  
   \bibitem{2aa}  T. Yu and J. H. Eberly, Finite-time disentanglement via spontaneous emission,  \href{https://journals.aps.org/prl/abstract/10.1103/PhysRevLett.93.140404} {Phys. Rev. Lett. \textbf{93}, 140404 (2004)}.
  
  \bibitem{2a} M. Yonac¸, T. Yu and J. H. Eberly, Sudden death of entanglement of two
Jaynes-Cummings atoms,  \href{http://iopscience.iop.org/article/10.1088/0953-4075/39/15/S09/meta;jsessionid=809C2B3A396BFD972D0919F83B543B2E.c4.iopscience.cld.iop.org} {J. Phys. B: At. Mol. Opt. Phys. \textbf{39},  S621 (2006)}.

 \bibitem{3} B. Bellomo, R. Lo Franco and G. Compagno, Non-Markovian effects on the dynamics of entanglement, \href{https://journals.aps.org/prl/abstract/10.1103/PhysRevLett.99.160502} {Phys. Rev. Lett. \textbf{99}, 160502 (2007)}.
  
 \bibitem{4} B. Bellomo, R. Lo Franco and G. Compagno, Entanglement dynamics of two independent qubits in environments with and without memory, \href{https://journals.aps.org/pra/abstract/10.1103/PhysRevA.77.032342} {Phys. Rev. A \textbf{77}, 032342 (2008)}.

 \bibitem{10} A. Orieux, A. D'Arrigo, G. Ferranti, R. Lo Franco, G. Benenti,
E. Paladino, G. Falci, F. Sciarrino and P. Mataloni, Experimental on-demand recovery of
entanglement by local operations within
non-Markovian dynamics, \href{https://www.nature.com/articles/srep08575} {Sci. Rep. \textbf{5}, 8575 (2015)}.



 
   \bibitem{11} Z.-X. Man, Y.-J. Xia and N. B. An, On conditions for atomic entanglement sudden
death in cavity QED, \href{http://iopscience.iop.org/article/10.1088/0953-4075/41/8/085503/meta} {J. Phys. B: At. Mol. Opt. Phys. \textbf{41},  085503 (2008)}.


\bibitem{11b} G. Argentieri, F. Benatti, R. Floreanini and U. Marzolino, Entangled identical particles and noise, \href{http://www.worldscientific.com/doi/abs/10.1142/S0219749911008210} { Int. J. Quantum Inform. \textbf{9}, 1745 (2011)}.

  \bibitem{9} S.-L. Chen, G.-Y. Chen and Y.-N. Chen, Increase of entanglement by local PT -symmetric operations, \href{https://journals.aps.org/pra/abstract/10.1103/PhysRevA.90.054301} {Phys. Rev. A \textbf{90}, 054301 (2014)}.
  

\bibitem{8}  T. F. Jordan, A. Shaji and E. C. G. Sudarshan, Entanglement increase from local interactions and not completely positive maps,  \href{https://journals.aps.org/pra/abstract/10.1103/PhysRevA.76.022102} {Phys. Rev. A \textbf{76}, 022102 (2007)}.  
  


 \bibitem{2b} M. Yonac¸, T. Yu and J. H. Eberly, Pairwise concurrence dynamics: a four-qubit model, 
  \href{http://iopscience.iop.org/article/10.1088/0953-4075/40/9/S02/meta} {J. Phys. B: At. Mol. Opt. Phys. \textbf{40}, S45  (2007)}.

  
 
  
  

  
  
\bibitem{6}  C. E. Lopez, G. Romero, F. Lastra, E. Solano, and J. C. Retamal, Sudden birth versus sudden death of entanglement in multipartite systems,  \href{https://journals.aps.org/prl/abstract/10.1103/PhysRevLett.101.080503} {Phys. Rev. Lett. \textbf{101}, 080503 (2008)}.
  
  
 \bibitem{7} C. E. Lopez, G. Romero and J. C. Retamal, Dynamics of entanglement transfer through multipartite dissipative systems,  \href{https://journals.aps.org/pra/abstract/10.1103/PhysRevA.81.062114} {Phys. Rev. A \textbf{81}, 062114 (2010)}.
 
 
 \bibitem{5}  B. Bellomo, G. Compagno, R. Lo Franco, A. Ridolfo and S. Savasta, Entanglement dynamics of two
independent cavity-embedded
quantum dots,  \href{http://iopscience.iop.org/article/10.1088/0031-8949/2011/T143/014004/meta} {Phys. Scr. \textbf{T143},  014004 (2011)}.
 
 
 

 
   

 \bibitem{11a} L. Li, J. Zou, Z. He, J.-G. Li, B. Shao and L.-A. Wu, New features of entanglement dynamics with initial system–bath correlations, \href{http://www.sciencedirect.com/science/article/pii/S0375960112000229} {Phys. Lett. A \textbf{376}, 913 (2012)}. 
 
 
 
  
  
  
\bibitem{20} M. A. Nielsen and I. L. Chuang, \textit{Quantum Computation and Quantum Information} (Cambridge University Press, Cambridge, England, 2000).    
  
    
    \bibitem{12} I. Sargolzahi and S. Y. Mirafzali, Structure of states for which each localized dynamics reduces to a localized subdynamics, \href{http://www.worldscientific.com/doi/abs/10.1142/S0219749917500435} {Int. J. Quantum Inform.
 \textbf{15}, 1750043 (2017)}. 

 \bibitem{13} I. Sargolzahi, Entanglement revival can occur only when the system-environment state is not a Markov state, \href{https://arxiv.org/abs/1706.00452} {arXiv:1706.00452}.

\bibitem{14} F. Mintert, A. R.R. Carvalho, M. Kus and A. Buchleitner, Measures and dynamics of entangled states,  \href{http://www.sciencedirect.com/science/article/pii/S0370157305002334} {Phys. Rep. \textbf{415}, 207 (2005)}.

\bibitem{15} W. K. Wootters, Entanglement of formation of an arbitrary state of two qubits,  \href{https://journals.aps.org/prl/abstract/10.1103/PhysRevLett.80.2245} {Phys.
Rev. Lett., \textbf{80}, 2245 (1998)}.

\bibitem{16} P. Rungta and C. M. Caves, Concurrence-based entanglement measures for isotropic states, 
 \href{https://journals.aps.org/pra/abstract/10.1103/PhysRevA.67.012307} {Phys. Rev. A \textbf{67}, 012307 (2003)}.

  
  \bibitem{17} G. Vidal, Entanglement monotones,  \href{http://www.tandfonline.com/doi/abs/10.1080/09500340008244048} {J. Mod. Opt. \textbf{47},  355 (2000)}.
   
   
  \bibitem{18} P. Hayden, R. Jozsa, D. Petz and A. Winter, Structure of states which satisfy strong subadditivity
of quantum entropy with equality,  \href{http://link.springer.com/article/10.1007/s00220-004-1049-z}{Commun. Math. Phys. {\bf 246}, 359 (2004)}.   
   
   
 \bibitem{19} F. Buscemi, Complete positivity, Markovianity, and the quantum data-processing inequality,
in the presence of initial system-environment correlations, \href{http://dx.doi.org/10.1103/PhysRevLett.113.140502} {Phys. Rev. Lett. {\bf 113}, 140502 (2014)}.  
    
   
  
  \bibitem{20a} R. Horodecki, P. Horodecki, M. Horodecki and K. Horodecki, Quantum entanglement, 
\href{https://journals.aps.org/rmp/abstract/10.1103/RevModPhys.81.865} {Rev.  Mod.  Phys. \textbf{81}, 865 (2009)}.
  

 
    
  
  
   \bibitem{20b} L.-A. Wu, Dressed qubits in nuclear spin baths,  \href{https://journals.aps.org/pra/abstract/10.1103/PhysRevA.81.044305} {Phys. Rev. A \textbf{81}, 044305 (2010)}.
  
  \bibitem{20c} O. Guhne and G. Toth, Entanglement detection,  \href{http://www.sciencedirect.com/science/article/pii/S0370157309000623} {Phys. Rep.
 \textbf{474}, 1  (2009)}.
  
  
 \bibitem{21a} H. Ollivier and W. H. Zurek, Quantum discord: A measure of the quantumness of correlations,  \href{https://journals.aps.org/prl/abstract/10.1103/PhysRevLett.88.017901} {Phys. Rev. Lett. \textbf{88}, 017901  (2001)}.  
  
  
   \bibitem{21}  C. A. Rodríguez-Rosario, K. Modi, A.-m. Kuah, A. Shaji and E. C. G. Sudarshan, Completely positive maps and classical correlations,  \href{http://iopscience.iop.org/article/10.1088/1751-8113/41/20/205301/meta} {J. Phys. A: Math. Theor. \textbf{41}, 205301 (2008)}.
   
   
  
       
     \bibitem{22} A. D'Arrigo, R. Lo Franco, G. Benenti,
E. Paladino and G. Falci, Recovering entanglement by local operations,  \href{http://www.sciencedirect.com/science/article/pii/S0003491614001997} {Ann. Phys. \textbf{350}, 211 (2014)}.
 
   
    \bibitem{23} R. Lo Franco, B. Bellomo, E. Andersson and G. Compagno, Revival of quantum correlations without system-environment back-action,  \href{https://journals.aps.org/pra/abstract/10.1103/PhysRevA.85.032318} {Phys. Rev. A \textbf{85}, 032318 (2012)}.
    
    
  

   

\end{thebibliography}
\end{document}